\documentclass[prl,twocolumn,preprintnumbers,superscriptaddress,showpacs]{revtex4-1}
\usepackage{graphicx}
\usepackage{amsmath}
\usepackage{amssymb}

\newcommand\sect[1]{\emph{#1} ---}
\newcommand{\p}{\ensuremath{{\mathbf{p}}}}
\newcommand{\q}{\ensuremath{{\mathbf{q}}}}
\renewcommand{\k}{\ensuremath{{\mathbf{k}}}}
\newcommand{\x}{\ensuremath{{\mathbf{x}}}}
\newcommand{\s}{\ensuremath{{\mathbf{s}}}}
\newcommand{\tr}{\ensuremath{{\mathbf{tr}}}}
\newcommand{\sgn}{\ensuremath{{\mathrm{sgn}}}}

\makeatletter
\newcommand*\wt[1]{\mathpalette\wthelper{#1}}
\newcommand*\wthelper[2]{%
        \hbox{\dimen@\accentfontxheight#1%
                \accentfontxheight#11.15\dimen@
                $\m@th#1\widetilde{#2}$%
                \accentfontxheight#1\dimen@
        }%
}
\newcommand*\accentfontxheight[1]{%
       \fontdimen5\ifx#1\displaystyle
                \textfont
        \else\ifx#1\textstyle
                \textfont
        \else\ifx#1\scriptstyle
                \scriptfont
        \else
                \scriptscriptfont
        \fi\fi\fi3
}
\makeatother

\begin{document}
%\begin{CJK}{UTF8}{gbsn}

\author{Jing-Yuan~Chen}
\affiliation{Stanford Institute for Theoretical Physics, Stanford University, Stanford, CA 94305}

\title{Static Magnetic Response of Non-Fermi Liquid Density}
%\date{April 2017}

\begin{abstract}
We consider the response of the density of a fermion ensemble to an applied weak static magnetic field. It is known that for non-interacting Fermi gas, this response is fully characterized by the Fermi volume and the Berry curvature on the Fermi surface. Here we show the same result holds for interacting fermions, including Fermi liquid and non-Fermi liquid, to all orders in perturbation theory. Our result relies only on the assumption of a well-defined Fermi surface and the general analytic properties of quantum field theory, and is completely model independent.
\end{abstract}

\maketitle
%\end{CJK}

\sect{Introduction}
Landau's theory of Fermi liquid \cite{landau1957theory, abrikosov1975methods} is the standard paradigm for a large class of interacting fermionic systems in $d>1$ spatial dimensions. A \emph{Fermi liquid} (FL) is characterized by two properties: The ground state is unique and has a well-defined Fermi surface (FS), and the low energy excitations are long-lived quasiparticles near the FS. These properties allow the low energy behaviors of the system to be captured by Landau's semi-classical picture, which has successfully explained numerous experimental phenomena over decades. On the other hand, there exist many interacting fermionic systems that do not fit into this picture. In particular, many are thought to be \emph{non-Fermi liquids} (NFL). In this letter, by NFL we mean an interacting fermionic system that still has well-defined FS in its ground state, but no long-lived quasiparticle at low energy. See Ref.~\cite{varma2002singular} for review. Very few physical properties have been concretely derived for NFLs. Dzyaloshinskii stressed \cite{dzyaloshinskii2003some} that Luttinger's Theorem \cite{Luttinger:1960zz},
\begin{align}
\rho = \frac{V_F+N V_{BZ}}{(2\pi)^d}, \label{LuttingerThm}
\end{align}
originally proven for FL, also holds in NFL. Here $\rho$ is the fermion density, $V_F$ is the volume enclosed by the FS (with species multiplicity taken into account), $V_{BZ}$ is the volume of the Brillouin zone (BZ), and $N$ is an integer corresponding to the number of filled bands; we have set $\hbar=1$. The same proof works for NFL as well as for FL because the theorem is a statement about the ground state, with no reference to the strange behavior of the low energy excitations in NFL. In this letter we show another ground state property shared by FL and NFL -- the generalized ``Luttinger's Theorem'' in the presence of a weak static magnetic field.

To present our result, we first briefly review the physics of Berry phase in fermions. Consider a non-interacting fermion described by the state $u^\alpha(\p)$, where $\p$ is a momentum in the BZ, and $\alpha$ runs over Bloch bands as well as spinor components. One can define the \emph{Berry connection} $a^i(\p) \equiv -i \, u^\dagger_\alpha(\p) \: \partial_p^i u^\alpha(\p)$ and the \emph{Berry curvature} $b^{ij}(\p) \equiv \partial_p^{i} a^{j}(\p)-\partial_p^{j} a^{i}(\p) = -2i \, \partial_p^{[i} u^\dagger_\alpha(\p) \: \partial_p^{j]} u^\alpha(\p)$; here $\partial_p^i \equiv \partial / \partial p_i$. They characterize the interference phase obtained by the state $u^\alpha$ when $\p$ is adiabatically changed under a weak force. The interference affects the average motion of a particle's wavepacket and gives rise to anomalous Hall effect; it also changes the Liouville phase space measure and induces an anomalous density. See Ref.~\cite{Xiao:2009rm} for review. In weak static magnetic field, the change of density at fixed chemical potential is given by
\begin{align}
\delta \rho = \sigma^{ij} \frac{B_{ij}}{2} + \mbox{(mag. dip.)}^{ij} \frac{B_{ij}}{2}
\label{AD}
\end{align}
($B_{ij}=B\epsilon_{ij}$ in $d=2$ and $B^k \epsilon_{ijk}$ in $d=3$, and the electric charge has been absorbed into the field); terms with $\partial_\x B$ or $B^2$ are neglected. The second term in \eqref{AD}, whose detailed expression is unimportant to us, is due to the fermions acquiring magnetic dipole potential energy, thereby shifting the FS. On the other hand, the first term, the anomalous density, does not involve a shift of the FS. The tensor
\begin{align}
\sigma^{ij} = \left\{ \begin{array}{ll}
\frac{\epsilon^{ij}}{2\pi} C + \frac{\epsilon^{ij}}{2\pi} \int_{FS}\frac{dp_k}{2\pi} \: a^k(\p), & \ d=2 \\[.2cm]
\frac{\epsilon^{ijk}}{2\pi} C_k + \frac{\epsilon^{ijk}}{2\pi} \int_{FS}\frac{(d^2p)_{mn}}{(2\pi)^2} \: p_k \: b^{mn}(\p), & \ d=3
\end{array} \right.
\label{HallCond}
\end{align}
is the Berry curvature contribution from both the filled bands and the FS \cite{Haldane:2004zz}; here $C$ in $d=2$ is an integer known as the total \emph{Chern number} of the filled bands, and analogously $C_k$ in $d=3$ is an integer combination of reciprocal lattice vectors. (We assume the FS does not intersect the boundary of our choice of BZ; if this happens, in $d=3$ there is an additional integral along the intersecting curve \cite{Haldane:2004zz}.)  Adding up \eqref{LuttingerThm} and \eqref{AD}, we can express
\begin{align}
\rho(B) = \frac{V_F(B)+N V_{BZ}}{(2\pi)^d} + \sigma^{ij} \frac{B_{ij}}{2},
\label{main}
\end{align}
where the shift of the FS due to the magnetic dipole moment is included in $V_F(B)$, the volume enclosed by the FS in the presence of the static magnetic field. Note that although $\p$ is no longer a good quantum number in the presence of the magnetic field, the FS is still resolvable to leading order in $B$ \cite{dunin1973two, Nguyen:2016itg}, and only becomes not well-defined if we care about precision to higher orders.

The fact that the conducting band contribution to \eqref{HallCond} can be written as FS integral, as opposed to Fermi sea integral, is a point emphasized by Haldane \cite{Haldane:2004zz}, for this makes the generalization towards interacting FL possible. Indeed, Ref.~\cite{Chen:2016fns} carried out a comprehensive study of Berry phase in interacting FL to all orders in perturbation theory. In the particular case of static magnetic field, \eqref{main} and \eqref{HallCond} remain valid, as long as we define $u^\alpha(\p)$ -- originally the single particle eigenstate needed to define $a^i$ and $b^{ij}$ -- by the eigenvector of the full Green's function of the interacting fermion at the FS. The diagrammatic derivation in Ref.~\cite{Chen:2016fns} has two limitations. It relies on the assumptions of FL, and tells nothing about NFL. It computes linear response, and hence does not capture the quantum oscillation \cite{luttinger1961theory,gor1962quantum} in FL.

In this letter, we show \eqref{main} and \eqref{HallCond} hold for NFLs to all orders in perturbation theory, with possible quantum oscillation taken into account. Our derivation relies only on the assumption of the presence of FS and the general analytic properties of quantum field theory. Although no comprehensive analysis of Berry phase in NFL can be made like that in FL, our result shows one particular response property of NFL can be unambiguously characterized by Berry phase.

It is important to ask whether our result is purely conceptual or has observable consequences. In fact the two terms in \eqref{main} separately have physical meanings in sudden approximation experiments. The FS in a weak static magnetic field can be resolved via scanning tunneling microscopy. The eigenvector $u^\alpha$ (referred to as \emph{pseudo-spin texture} when two-component) near the FS can be approximated by that in the absence of magnetic field, and can be resolved by scanning tunneling microscopy or photoemission spectroscopy. Although we are not making experimental proposal in this letter, our result is measurable at least in principle.

As our result is simple and experimentally verifiable, and its derivation relies only on general first principles, it poses a non-trivial statement on future phenomenological theory proposals about strongly interacting fermionic systems.

We first present our derivation in static homogeneous magnetic field by generalizing Luttinger's original proof to his theorem \cite{Luttinger:1960zz, abrikosov1975methods, dzyaloshinskii2003some}. Then we consider the inhomogeneous case. Finally we make concluding remarks.

\sect{In Homogeneous Magnetic Field}
Consider fermions under an external static vector potential $A_i(\x)$. Let $iG^\alpha_{\ \beta}(\omega, \x_1, \x_2) \equiv \int_{-\infty}^\infty dt \: e^{i\omega t} \langle T \: \psi^\alpha(t, \x_1) \: \psi^\dagger_\beta(0, \x_2) \rangle$ be the full (time-ordered) Green's function; there is no spatial translational invariance due to the presence of the external field. Let $\s\equiv \x_1-\x_2$ and $\x \equiv (\x_1+\x_2)/2$, we also denote $G^\alpha_{\ \beta}(\omega, \s; \x) \equiv G^\alpha_{\ \beta}(\omega, \x_1, \x_2)$. The fermion density at $\x$ is given by
\begin{align}
i\rho(\x) &= -\int_\omega e^{i\omega 0^+} i\delta^\beta_{\ \alpha} \ iG^\alpha_{\ \beta}(\omega, \x, \x) \nonumber \\[.1cm]
&= \int_{\omega, \s} e^{i\omega 0^+} \: \partial_{\omega}(G_0^{-1})^\beta_{\ \alpha} (\omega, -\s; \x) \: G^\alpha_{\ \beta}(\omega, \s; \x)
\end{align}
where $(G_0^{-1})^\beta_{\ \alpha}(\omega, \x_2, \x_1) \equiv \langle \x_2, \beta | \left( \omega - H_0 \right) | \x_1, \alpha \rangle$ is the inverse non-interacting Green's function; its $\omega$ derivative is just $\delta^\beta_\alpha \delta^d(\s)$. In this letter we abbreviate $\int_t \equiv \int dt, \: \int_\s \equiv \int d^d s$ while $\int_\omega \equiv \int d\omega/2\pi, \: \int_\p \equiv \int d^d p/(2\pi)^d$; as usual, the $\omega$ integral is along the complex contour $\mathrm{Re}\, \omega + i0^+ \sgn\, \mathrm{Re}\, \omega$. Using $G^{-1} = G_0^{-1}-\Sigma$ where $\Sigma$ is the self energy, we have
\begin{align}
\rho(\x) =& \ i\int_{\omega, \s} e^{i\omega 0^+} \: (G^{-1})^\beta_{\ \alpha} (\omega, -\s; \x) \: \partial_{\omega} G^\alpha_{\ \beta}(\omega, \s; \x) \nonumber \\[.1cm]
& + \ i\int_{\omega, \s} e^{i\omega 0^+} \: \Sigma^\beta_{\ \alpha} (\omega, -\s; \x) \: \partial_{\omega} G^\alpha_{\ \beta}(\omega, \s; \x).
\label{starting_point}
\end{align}
Note that the integration of $\omega$ by parts has no boundary term due to the general properties that $\Sigma\rightarrow const.$ and $G\rightarrow \mathbf{1}/\omega$ as $\omega\rightarrow \pm \infty$.

Similar to the usual case without external field, the second line of \eqref{starting_point} vanishes in homogeneous magnetic field, due to the following. Diagrammatically, under functional variation the self energy satisfies $\delta \Sigma_I / \delta G_{\wt J} = -i \wt{V}_{IJ}$ where we collectively denoted $I=(\omega, (\beta, \x_2), (\alpha, \x_1))$, and $\wt I=(\omega, (\alpha, \x_1), (\beta, \x_2))$ as opposed to $I$. Here $\wt V_{I J}$ is the usual 2-particle irreducible scattering blob, which is symmetric between $I$ and $J$. Thus, there exists some functional $\Phi$ such that $\Sigma_I = \delta \Phi/\delta G_{\wt I}$. Therefore, if the magnetic field is homogeneous, the second line of \eqref{starting_point} is equal to $\mathcal{V}^{-1} i\int_I (\delta \Phi/\delta G_{\wt I}) \partial_\omega G_{\wt I}$ (where $\mathcal{V}$ is the spatial volume) and vanishes as a total $\omega$ derivative.

We are left with the first line of \eqref{starting_point}, which, under Fourier transformation from $\s$ to $\p$, reads
\begin{align}
\rho(\x) &= i\int_{\omega, \p} e^{i\omega 0^+} \: (G^{-1})^\beta_{\ \alpha} (\omega, \p; \x) \: \partial_{\omega} G^\alpha_{\ \beta}(\omega, \p; \x).
\label{mid_point}
\end{align}
In the usual case without external field, the integrand would then be written as $\partial_\omega \ln \det G (\omega, \p)$. But now one must be careful that, due to the $\x$ dependence, $(G^{-1})(\omega, \p; \x) \equiv \int_\s e^{-i\p\cdot(-\s)} (G^{-1})(\omega, -\s; \x)$ is no longer the same as $\left(G(\omega, \p; \x) \right)^{-1}$. We now find their difference. By definition (matrix multiplication and the presence of $\omega$ are understood in the equations below)
\begin{align}
\delta^d(\x_1-\x_3) \: \mathbf{1} = \int_{\x_2} (G^{-1})(\x_1, \x_2) \: G(\x_2, \x_3).
\end{align}
Fourier transforming to momentum space, and using $G(\p+\k/2, \p-\k/2) = \int_\k e^{-i\k\cdot\x}G(\p; \x) $, we have
\begin{align}
(2\pi)^d \delta^d(\q) \: \mathbf{1} =& \int_{\k, \x, \x'} e^{-i(\q-\k)\cdot\x - i \k\cdot\x'} \nonumber \\[.1cm]
 & (G^{-1})(\p+\k/2; \x) \: G(\p+\k/2-\q/2; \x').
 \label{GinvFS}
\end{align}
Now we perform gradient expansion. The expansion of $G^{-1}$ in $\partial_\x$ is equivalent to the expansion of $G$ in $\q-\k$, due to the $e^{-i(\q-\k)\cdot \x}$ factor; likewise, the expansion of $G$ in $\partial_{\x'}$ is equivalent to the expansion of $G^{-1}$ in $\k$. To first order, we find
\begin{align}
(G^{-1})(\p; \x) =& \left( \mathbf{1} - \frac{i}{2} \ \partial_\x \left( G(\p; \x)\right)^{-1} \cdot \partial_\p G(\p; \x) \right. \nonumber \\[.1cm]
& \left. \ \ \ \ + \ \frac{i}{2} \ \partial_\p \left( G(\p; \x)\right)^{-1} \cdot \partial_\x G(\p; \x) \right) \nonumber \\[.1cm]
& \ \ \left(G(\p; \x)\right)^{-1}
\end{align}
to be substituted into \eqref{mid_point}.

In our case, the $\x$ dependence comes from $A_i(\x)$. If $B_{ij}$ is homogenous, then homogeneity and gauge covariance requires $G(\p; \x)$ to depend only on the combination $p_i-A_i(\x)=p_i+B_{ij}x^j/2$ (in symmetric gauge) \cite{gor1962quantum}. Thus, 
\begin{align}
\rho(\x) =& \ i\int_{\omega, \p} e^{i\omega 0^+} \: \partial_\omega \ln\det G \nonumber \\[.1cm]
& + \frac{B_{ij}}{2} \int_{\omega, \p} e^{i\omega 0^+} \tr \left(\partial_p^i G^{-1} \, G \, \partial_p^j G^{-1} \, G \, \partial_\omega G^{-1} G\right).
\label{FS_Berry}
\end{align}
where $G$ means $G^\alpha_{\ \beta}(\omega, \p; \x)$ and $G^{-1}$ means $((G(\omega, \p; \x))^{-1})^\gamma_{\ \delta}$. The two terms in \eqref{FS_Berry} will respectively give rise to the two terms in \eqref{main}.

The first term of \eqref{FS_Berry} is evaluated the in same way as the usual case without external field. The integrand can be written as $\sum_n \partial_\omega \ln \lambda_n$ where $\lambda_n$ is the $n$th eigenvalue of $G^\alpha_{\ \beta}$. Due to the $e^{i\omega 0^+}$ factor and the fact that $G$ is analytic without zeroes when $\omega$ is not real, the contour of the $\omega$ integral can be deformed into the contour enclosing around the negative real axis. Thus, for given $n$, $\p$ and $\x$, the $\omega$ integral counts the winding number (multiplied by $i$) of the complex phase of $\lambda_n(\omega, \p; \x)$ as $\omega$ winds around the negative real axis. From the analytic properties $G(\omega^\ast, \p; \x) = G(\omega, \p; \x)^\dagger$ and $G\rightarrow \mathbf{1}/\omega$ as $\omega\rightarrow -\infty$, the phase winding is even (odd) if $\lambda_n(0, \p; \x)<0 \ (>0)$. In the absence of external field, the analytic property $\sgn\, \mathrm{Im}\, \lambda_n=-\sgn\, \mathrm{Im}\, \omega$ further demands the even (odd) phase winding to be $0 \ (-1)$; although this analytic property is lost due to the external field, the phase winding being $0 \ (-1)$ should still hold as we expect the weak magnetic field not to dramatically change the phase of the system. Finally, integrating over $\p$ and summing over $n$, we obtain the first term of \eqref{main}, with the FS given by values of $\p$ where $\lambda_n(0, \p; \x)=0$.

The second term of \eqref{FS_Berry} gives rise to the Berry curvature \cite{Haldane_talk, Chen:2016fns}. Note we can antisymmetrize between $\partial_p^i, \partial_p^j, \partial_\omega$ in the integrand. The integrand, formally being a Wess-Zumino-Witten integrand, is a closed differential form, so it must be locally exact. More explicitly, diagonalizing $G=U\Lambda U^{-1}$, one can show (denoting $\partial_p^0\equiv -\partial_\omega$)
\begin{align}
& \ \tr \left(\partial_p^{[i} G^{-1} \, G \, \partial_p^{j} G^{-1} \, G \, \partial_p^{0]} G^{-1} G\right) \nonumber \\[.1cm]
=& \ 6\, \partial_p^{[i} \: \tr\left( U^{-1} \partial_p^{j} U \ \partial_p^{0]} \ln \Lambda\right) \nonumber \\[.1cm]
& -3\, \partial_p^{[i} \: \tr\left(\Lambda^{-1} U^{-1} \partial_p^j U \, \Lambda \, U^{-1} \partial_p^{0]} U\right)
\label{WZW_local_exact}
\end{align}
via brute force computation. There are two kinds of discontinuities making \eqref{WZW_local_exact} globally non-exact. First, as before, across the FS some eigenvalue(s) $\lambda_n$ in $\Lambda$ changes its phase winding. Second, $U$ may not be continuously defined over the entire BZ, and is subjected to gauge transformation $U\rightarrow UW$ across some $d-1$ dimensional patch boundary $\mathcal{M}$ in the BZ \cite{Thouless:1982zz}, where the transition function $W$ is a unitary (block) diagonal matrix that commutes with $\Lambda$. (In general $U$ has $\omega$ dependence, but we can make $\mathcal{M}$ and $W$ independent of $\omega$, because the $\omega$ dependence of $U$ has no compact boundary condition). Neither kind of discontinuity comes up in the second term of \eqref{WZW_local_exact}, because its trace does not involve $\partial_\omega \Lambda$ and is invariant under $U\rightarrow UW$. Hence this term vanishes as a total derivative upon the $\omega, \p$ integrals. Both kinds of discontinuity appear in the first term, giving rise to
\begin{align}
\sigma^{ij} = \sum_n \int_{\omega, \p} e^{i\omega 0^+} 2 \: \partial_p^{[i} \left( (u^{-1}_\alpha)_n \partial_p^{j]} u^\alpha_n \ \partial_\omega \ln \lambda_n \right)
\end{align}
with the contour of the $\omega$ integral again deformed to enclose around the negative real axis. The discontinuity in the phase winding of $\lambda_n$, placing branch cut along $\lambda_n>0$, is picked up across the FS, with discontinuity $\Delta \left( \partial_\omega \ln \lambda_n \right) = 2\pi i \delta(\omega)$. The discontinuity in $u_n$ is picked up as $p_i$ runs across $\mathcal{M}$ when $\lambda_n$ has phase winding $-1$ (this can happen in the filled bands or in the ``Fermi sea'' of the conducting band(s)); the discontinuity across $\mathcal{M}$ is $\Delta \left( (u^{-1}_\alpha)_n \partial_p^{j} u^\alpha_n \right) = w_n^\ast \partial_p^j w_n$. Thus we find
\begin{align}
\sigma^{ij} =& \sum_{n\in\mbox{\scriptsize cond.}} 2\left(i \int_{\p \in FS} \mathrm{n}_{FS}^{[i} \: (u_\alpha^{-1})_n \partial_p^{j]} u^\alpha_n \right. \nonumber \\[.1cm]
& \hspace{1.5cm} \left. +\: (-i) \int_{\p \in \mathcal{M}\cap \mbox{\scriptsize Fermi sea}} \mathrm{n}_{\mathcal{M}}^{[i} \: w_n^\ast \partial_p^{j]} w_n \right) \nonumber \\[.2cm]
& + \: \sum_{n\in\mbox{\scriptsize filled}} 2\: (-i) \int_{\p \in \mathcal{M}} \mathrm{n}_{\mathcal{M}}^{[i} \: w_n^\ast \partial_p^{j]} w_n
\end{align}
where $\mathrm{n}_{FS}^i(\p)$ and $\mathrm{n}_{\mathcal{M}}^i(\p)$ are unit normal vectors to the FS and to $\mathcal{M}$ respectively. Clearly the filled bands contribute the quantized values $C$ or $C_k$ in \eqref{HallCond} \cite{Thouless:1982zz}, in the same way as in usual gapped Chern insulators. For the conducting band(s), $u$ in the first term is evaluated at $\omega=0$, at which $G^\alpha_{\ \beta}$ becomes Hermitian and hence $U^{-1} = U^\dagger$. The first term can therefore be recognized as $-2\int_{FS} \mathrm{n}_{FS}^{[i} a^{j]}$. Note that the Berry connection is well-defined by the eigenvectors of $G$, regardless of the existence of quasiparticles, which relies on the form of the eigenvalues of $G$. In $d=2$, we can always choose $\mathcal{M}$ to lie outside of the Fermi sea so that the second term vanishes, and the first term is equal to the $d=2$ FS term in \eqref{HallCond}. In $d=3$, if the Berry curvature is exact on the FS (no net Weyl node enclosed in each FS component), $\mathcal{M}$ can still be chosen to lie outside of the Fermi sea, and the first term reduces to the $d=3$ FS term in \eqref{HallCond} via integration by parts \cite{Haldane:2004zz, Haldane_talk}. If the Berry curvature is non-exact on the FS (net Weyl node enclosed in some FS components), $\mathcal{M}$ must intersect the FS, and the integration by parts involves an extra term due to the discontinuity $\Delta a^j$ on $\mathcal{M}\cap FS$; but this extra term from integration by parts just cancels the second term which is now non-vanishing by itself. (The derivation of \eqref{HallCond} here is stronger than that in \cite{Chen:2016fns}; the latter only showed the chemical potential dependence on both sides agree.)

We have thus proven \eqref{main} and \eqref{HallCond} for interacting fermions under very general assumptions. A linear response computation would also give, among other terms, the second line of \eqref{FS_Berry}, but those other terms are not organized into a compact form like the first line of \eqref{FS_Berry}. This makes the linear response result hard to interpret in certain NFLs (see below); moreover, even in FLs, quantum oscillation, being a non-analytic effect in $A_i$, is missed from linear response. On contrary, our present method can concisely conclude on these issues. In particular, under quantum oscillation, $\rho(B)$ and $V_F(B)$ oscillate together, while their deviation is always characterized by the Berry curvature from the FS and the filled bands.

\sect{In Inhomogeneous Magnetic Field}
When the external static magnetic field is slightly inhomogeneous with a wave vector $\q\sim -i\partial_\x$ much smaller than the FS size, we expect \eqref{main} to still hold, with $\rho(B(\x))$ the local density and $V_F(B(\x))$ the volume enclosed by the FS measured near $\x$ (technically given by $\lambda_n(0, \p; \x)=0$). Although the derivation presented above relied on a homogeneous magnetic field, the slight inhomogeneity should only give corrections of order $\partial_\x B$, which we shall ignore in the spirit of gradient expansion.

We can also see this via a linear response computation like that in Ref.~\cite{Chen:2016fns}, though this method is less conclusive. We just sketch the procedure here. When the external field is static, no excitation occurs; diagrammatically this corresponds to no contribution from Cutkosky cuts, and we can safely perform $\q$ expansion in the diagrams, and keep $\q$ to first order. Let $i\left(\underline\Gamma^\mu\right)^\alpha_{\ \beta}(\omega, \p)$ be the full electromagnetic vertex coupling a fermion with energy and momentum $(\omega, \p)$ to $A_\mu(q^0, \q)$ in the limit $q^0=0, \ \q\rightarrow 0$. It is related to $G$ via the gapless Ward-Takahashi identity \cite{abrikosov1975methods, varma2002singular}
\begin{align}
\underline\Gamma^i = -\partial_p^i G^{-1}, \ \ \ \ \underline\Gamma^0 = \partial_\omega G^{-1} - G^{-1} \: \partial_F G \: G^{-1}
\end{align}
where $\partial_F \equiv \partial/\partial \mu - \partial / \partial \omega$ ($\mu$ is the chemical potential) captures the physical effect of changing the ground state (the subtraction of $\partial/\partial \omega$ is because we always define $\omega=0$ on the FS). Clearly $\partial_F G$ behaves differently in FL and NFL; however, it can always be interpreted as the shift of the FS, by an amount depending also on other quantities contracted with it. Therefore, all terms involving $\partial_F G$ eventually shifts $V_F$ to $V_F(B)$ in \eqref{main}. (The shift of FS has finite gradient in e.g. marginal NFLs \cite{varma1989phenomenology, varma2002singular}, but has diverging gradient in some critical NFLs with scaling exponents in certain ranges \cite{senthil2008critical}. In the latter case, this linear response method is inconclusive.) The remaining terms have the same form for FL and NFL in terms of the full Green's function $G$ (including its $\omega, \p$ derivatives) and bare interaction vertices. These remaining terms sum up in the density to \cite{Chen:2016fns}
\begin{align}
& \frac{B_{ij}(\q)}{2} \int_{\omega, \p} e^{i\omega 0^+} \tr \left(\partial_p^{[i} G^{-1} \, G \, \partial_p^{j]} G^{-1} \, G \, \partial_\omega G^{-1} G\right) \nonumber \\[.2cm]
& + \frac{B_{ij}(\q)}{2} \: \sigma_Y^{ij0}.
\end{align}
The first line just gives rise to the $\sigma^{ij}$ term in \eqref{main} as in the homogeneous case. In the second line, $\sigma_Y^{ij0}$ is a complicated sum of interaction diagrams as defined in Appendix D of Ref.~\cite{Chen:2016fns}, and is shown to vanish there. Therefore, the linear response computation suggests \eqref{main} holds in slightly inhomogeneous magnetic field.

\sect{Conclusion}
We presented a very general derivation of \eqref{main} for interacting fermionic systems, including FLs and NFLs, in weak static homogeneous magnetic field. We argued the same should hold when there is slight inhomogeneity. We are unaware of a non-perturbative derivation of our result like Oshikawa's proof to Luttinger's Theorem \cite{Oshikawa:2000lrt}; in particular, Oshikawa's proof relies crucially on the notion of quasiparticles, and it is unclear how to extend it to NFLs (even in the absence of external field). We have not considered impurities in our derivation. But as our result is about ground state density, we expect impurities not to spoil our result, as much as they do not spoil Luttinger's Theorem.

Our analysis can be easily adapted to compute the steady current induced by a static magnetic field. As expected, the steady current is just the chiral magnetic effect protected by the $U(1)$ chiral anomaly \cite{Son:2009tf}. For topological reasons, the chiral magnetic effect is always zero for fermions on a lattice when in equilibrium \cite{Basar:2013iaa, [][. See {\it Chp. 2.5.3.}]Chen:2016lra}.

\acknowledgments
The author thanks F.~D.~M.~Haldane, S.~A.~Kivelson and D.~T.~Son for helpful discussions. The author is funded by the Gordon and Betty Moore Foundation's EPiQS Initiative through Grant GBMF4302.

\bibliography{NFL_Berry}

\end{document}